# Superconductor-Insulator Transition in a Disordered Electronic System

Nandini Trivedi*,[1] Richard T. Scalettar,[2] and Mohit Randeria*[1]

[1] *Materials Science Division 223, Argonne National Laboratory, Argonne, IL 60439*
[2] *Department of Physics, University of California, Davis, CA 95616*

We study an electronic model of a 2D superconductor with onsite randomness using Quantum Monte Carlo simulations. The superfluid density is used to track the destruction of superconductivity in the ground state with increasing disorder. The non-superconducting state is identified as an insulator from the temperature dependence of its d.c. resistivity. The value of $\sigma_{\rm dc}$ at the superconductor-insulator transition appears to be non-universal.

The problem of the effect of strong disorder on superconductivity and of the resulting superconductor(SC)-insulator(I) transition in low dimensional systems has been studied experimentally for a number of years [1]. Theoretically, the problem is challenging because of the complicated interplay between interactions and disorder [2]. Within mean-field theory [3] superconductivity appears to persist essentially all the way to the site localized limit due to an inadequate description of the disorder-induced fluctuations of the local order parameter. Much of the recent theoretical effort has thus focused on the dirty boson problem [4] which is expected to capture the essential physics of these fluctuations. The boson models which describe the universal critical properties *at* the SC-I transition are also more amenable to analytical [4] and numerical [5,6] studies. However, if one is interested in characterizing the phases, and studying the question of a possible "universal" conductance at the SC-I transition, one has to go back to a description in terms of the electronic degrees of freedom.

As a first step in this direction we use Quantum Monte Carlo (QMC) simulations to study the simplest fermionic problem which can have superconducting, insulating, and (possibly) metallic phases. This is the attractive Hubbard model with an onsite random potential

$$H = -t \sum_{\langle ij\rangle\sigma} (c^\dagger_{i\sigma} c_{j\sigma} + c^\dagger_{j\sigma} c_{i\sigma}) \\ - \sum_{i\sigma}(\mu - v_i)n_{i\sigma} - |U|\sum_i n_{i\uparrow}n_{i\downarrow}. \quad (1)$$

We set $t=1$ and measure all energies in units of $t$. Here $c_{i\sigma}$ is a fermion destruction operator at site $i$ with spin $\sigma$, $n_{i\sigma} = c^\dagger_{i\sigma}c_{i\sigma}$, and the chemical potential $\mu$ fixes the average density $\langle n \rangle$. The site energies $v_i$ are independent random variables with a uniform distribution over $[-V,V]$. The lattice sum $\langle ij \rangle$ is over near neighbor sites on a two dimensional square lattice. Note that this model focuses on the localization induced by the disorder; it does not, however, incorporate the disorder-dependence of the effective electron-electron interaction [2].

Our main results can be summarized as follows:
1) The transverse current-current correlation is used to compute the superfluid stiffness $D_s$. At low temperatures, $D_s$ decreases with increasing disorder $V$. Beyond a critical $V_c$ the system becomes non-superconducting.
2) For $V > V_c$ the system is an insulator, as seen from the $T$-dependence of its dc resistivity $\rho_{\rm dc}$: $d\rho_{\rm dc}/dT < 0$. No evidence for a metallic phase is seen in 2D. A simple analytic continuation method, argued to be valid for disordered systems, is used to estimate $\rho_{\rm dc}$ for the first time using such QMC methods.
3) The dc resistivity as a function of $T$ for various disorder strengths is also used to independently estimate the critical $V_c$ for the SC-I transition. This estimate is in good agreement with the one obtained from $D_s$.
4) $\rho_{dc}$ at $V = V_c$ depends on interaction strength $U$ and appears to be non-universal.

Before turning to the results of the interacting, disordered problem, we briefly summarize various limiting cases. In the noninteracting limit ($|U| = 0$) one obtains the Anderson localization problem, which is known to be insulating for all $V$ in 2D [2]. In the absence of disorder ($V=0$) one obtains the attractive Hubbard model. Off half-filling, $\langle n \rangle \neq 1$, the ground state has long range SC order and a finite temperature Kosterlitz-Thouless transition [7]. QMC simulations have played an important role in the study of the non-random model, especially in its anomalous normal state behavior [8]. The same algorithms [9], which are still free of the fermion sign problem, are applied here to the model (1).

We shall focus here mainly on various quantities that can be obtained from the current-current correlation function $\Lambda$ [10]. The (paramagnetic piece of the) current operator is defined as

$$j_x(\mathbf{l}\,\tau) = e^{H\tau}\left[it\sum_\sigma (c^\dagger_{\mathbf{l}+\hat x,\sigma}c_{\mathbf{l},\sigma} - c^\dagger_{\mathbf{l},\sigma}c_{\mathbf{l}+\hat x,\sigma})\right]e^{-H\tau}. \quad (2)$$

The Fourier transform of the impurity averaged $\Lambda_{xx}$ is then given by

$$\Lambda_{xx}(\mathbf{q}; i\omega_n) = \sum_{\mathbf{l}} \int_0^\beta d\tau \langle j_x(\mathbf{l}, \tau) j_x(0,0) \rangle e^{i\mathbf{q}\cdot\mathbf{l}} e^{-i\omega_n \tau}, \quad (3)$$

where $\omega_n = 2n\pi/\beta$, and $\langle \ldots \rangle$ denotes a thermal average at a temperature $T = \beta^{-1}$ for a given realization of disorder *and* an average over an ensemble of such realizations.

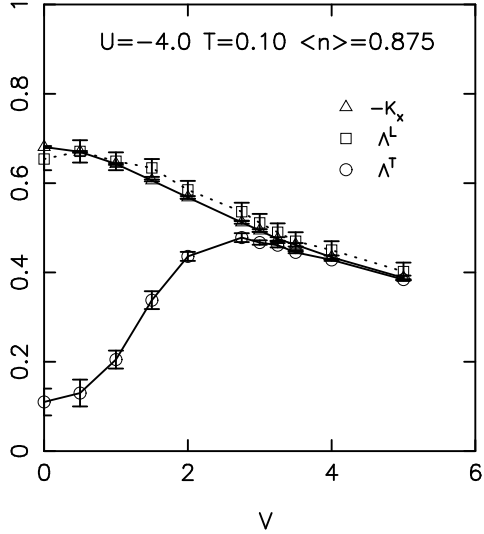

FIG. 1(a). The kinetic energy, $-K_x$, and transverse and longitudinal current-current correlation functions, $\Lambda^T$ and $\Lambda^L$, are shown as a function of disorder $V$. $\Lambda^L$ tracks $-K_x$ as required by gauge invariance. The difference of $\Lambda^T$ from $K_x$ signals the formation of a SC state at weak disorder (see Eq. (5)).

Gauge invariance requires that the longitudinal part of $\Lambda$ satisfy the equality [10,11]

$$\Lambda^L \equiv \lim_{q_x \to 0} \Lambda_{xx}(q_x, q_y = 0; i\omega_n = 0) = -K_x, \quad (4)$$

where $K_x = \langle -t \sum_\sigma (c^\dagger_{\mathbf{l}+\hat{x},\sigma} c_{\mathbf{l},\sigma} + c^\dagger_{\mathbf{l},\sigma} c_{\mathbf{l}+\hat{x},\sigma}) \rangle$, the kinetic energy in the $x$ direction, represents the diamagnetic part of the response. We have verified this equality as a non-trivial check on our numerics; see Fig. 1(a).

To study the destruction of SC with increasing disorder we look at the superfluid stiffness $D_s$ obtained from the transverse current-current correlation function [10,11]

$$\Lambda^T \equiv \lim_{q_y \to 0} \Lambda_{xx}(q_x = 0, q_y; i\omega_n = 0) \quad (5)$$
$$D_s = \pi[-K_x - \Lambda^T]$$

Results [12] for $\Lambda^T, \Lambda^L$ and $-K_x$ are plotted in Fig. 1(a) for $U = -4$, $T = 0.10$, and $\langle n \rangle = 0.875$ as a function of $V$. $\Lambda^T$ is estimated by using a linear extrapolation of the two smallest $q_y$ values. In Fig. 1(b) we plot $D_s$, which decreases monotonically with disorder. There is a critical value $V_c$ beyond which $D_s = 0$ and the system becomes non-superconducting.

Another quantity of interest is the $T = 0$ charge stiffness [10] $D = \pi[-K_x - \lim_{\omega \to 0} \text{Re}\,\Lambda_{xx}(\mathbf{q} = 0; \omega + i0^+)]$, which is the strength of delta function $\delta(\omega)$ in the optical conductivity. We find that $D$ is indeed non-zero in the SC state and, within the accuracy of our numerics, we always find $D \simeq D_s$, as shown in Fig. 1(b). However, $D$ is not useful in characterizing the non-superconducting state for $V > V_c$. In contrast to non-random systems, where $D$ can be used to distinguish a metal ($D \neq 0$) from an insulator ($D = 0$), for a disordered system $D \equiv 0$ for any non-superconducting state (since a dirty metal does not have a $\delta$-function in its $T = 0$ optical conductivity).

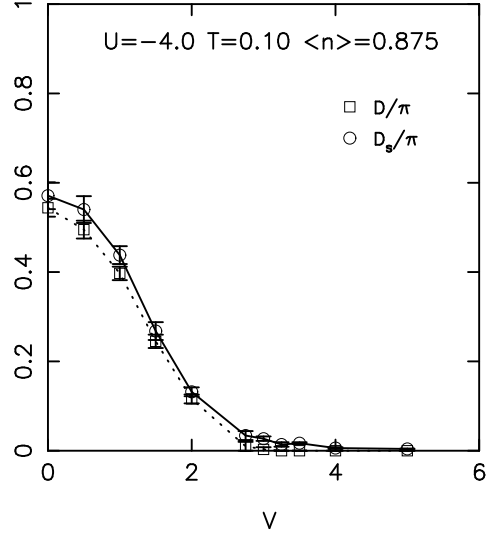

FIG. 1(b). The superfluid density $D_s$ and charge stiffness $D$, as a function of disorder $V$. $D_s = D$ for all $V$.

We must therefore turn to the dc conductivity to distinguish a metal from an insulator. This is given by $\sigma_{\text{dc}} = \lim_{\omega \to 0} \text{Im}\Lambda(\mathbf{q} = 0; \omega)/\omega$, where $\Lambda(\mathbf{q}; \omega + i0^+) = \text{Re}\Lambda(\mathbf{q}; \omega) + i\text{Im}\Lambda(\mathbf{q}; \omega)$, and we drop the $xx$ subscripts for simplicity. Using the fluctuation-dissipation theorem we obtain

$$\Lambda_{xx}(\mathbf{q}; \tau) = \int_{-\infty}^{+\infty} \frac{d\omega}{\pi} \frac{\exp(-\omega\tau)}{[1 - \exp(-\beta\omega)]} \text{Im}\Lambda_{xx}(\mathbf{q}; \omega), \quad (6)$$

valid for $0 \leq \tau \leq \beta$. To obtain $\text{Im}\Lambda$ from $\Lambda(\mathbf{q}; \tau)$, which is computed in the QMC, requires a numerical inversion of the Laplace transform. We will instead use a technique [8] which is valid for $T \ll \Omega$, where $\Omega$ is the scale on which $\text{Im}\Lambda$ deviates from its low frequency asymptotic behavior ($\text{Im}\Lambda \simeq \omega \sigma_{\text{dc}}$). Provided $T \ll \Omega$, Eq. (6) can be simplified to

$$\Lambda_{xx}(\mathbf{q} = 0; \tau = \beta/2) = \pi \sigma_{\text{dc}}/\beta^2, \quad (7)$$

which yields the dc conductivity. We note that this simplification may *not* be valid for non-random systems: e.g., for a Fermi liquid the scale $\Omega \simeq 1/\tau_{e-e} \sim N(0)T^2$ so one can never satisfy $T \ll \Omega$ at low $T$. However, for the highly disordered state that we study, we expect the scale $\Omega$ to be set by the disorder $V$ and to be $T$-independent, so that Eq. (7) is valid. We will present below additional consistency checks of this approximation.

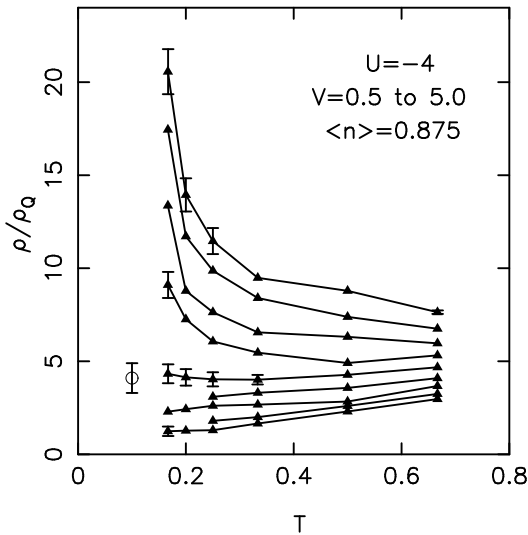

FIG. 2(a). $\rho_{dc}$ using Eq. (7) as function of temperature $T$ for various disorder strengths $V$. At weak $V$, $d\rho/dT > 0$, but for large V $d\rho/dT < 0$ (insulating). The point at $T = 0.10$ is obtained using Eq. (8) for $V \simeq V_c$.

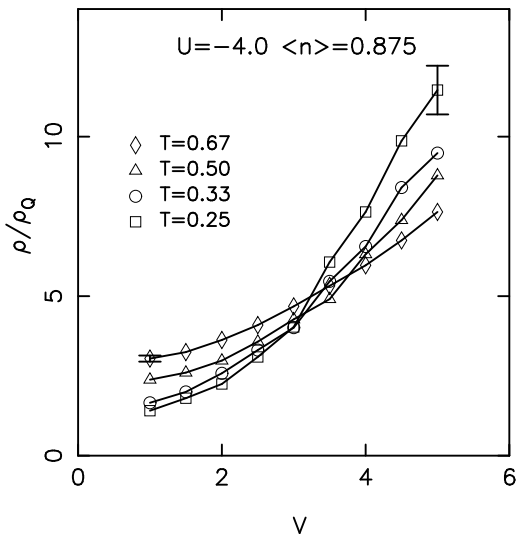

FIG. 2(b). $\rho_{dc}$ as a function of disorder $V$ for various temperatures $T$. Representative error bars are shown.

In Fig. 2(a) we plot the dc resistivity $\rho_{dc} = 1/\sigma_{dc}$ as a function of temperature for various degrees of disorder. We use units where $e^2 = \hbar = 1$ so that the quantum of resistivity $\rho_Q = h/(4e^2) = \pi/2$. For small disorder we see that $\rho_{dc}$ decreases with lowering $T$; that the underlying state is a SC can only really be seen from the superfluid density $D_s$. With increasing disorder the $T$-dependence is altered qualitatively: for large $V$ we see that $d\rho_{dc}/dT < 0$, strongly suggestive of insulating behavior ($\rho_{dc} = \infty$ at $T = 0$). To see precisely where this transition takes place it is useful to replot the data of Fig. 2(a) as $\rho_{dc}$ versus the strength of the disorder $V$, with different curves corresponding to different temperatures. This is done in Fig. 2(b); from the crossing point of the various curves we estimate the critical disorder $V_c$ separating the SC from the insulator. This crossing plot is not a consequence of any scaling ansatz, but simply follows from the monotonic behavior of $\rho_{dc}$ with $T$ which we find in the QMC. These curves are remarkably similar to those found in the experimental literature [1], and represent the first QMC calculations of the resistivity in a disordered, interacting fermi system.

An independent estimate of $\sigma_{dc}$ may be obtained as follows: We use the Matsubara conductivity

$$\sigma(i\omega_n) = [-K_x - \Lambda_{xx}(\mathbf{q}, i\omega_n)]/\omega_n \qquad (8)$$

obtained from the simulation and fit it to a Drude form [15] $\sigma_D(i\omega_n) = \sigma_{dc}/[1 + |\omega_n|\tau]$, which upon analytic continuation yields $\sigma_D(\omega + i0^+) = \sigma_{dc}/[1 - i\omega\tau]$. The simple Drude form is expected to be valid for a metallic system at low temperatures, where the $\omega$ and $T$-independent impurity scattering rate $1/\tau$ dominates any inelastic contribution due to the interactions. We thus use this procedure to extract the low temperature $\rho_{dc} = \sigma_{dc}^{-1}$ for $V \simeq V_c$. The value obtained at $T = 0.10$ is shown in Fig. 2(a), and is in excellent agreement with the estimates obtained from Eq. (7) at higher temperatures. $\sigma(i\omega_n)$ could be well fit by a Drude form only for $V \simeq V_c$, suggesting the absence of an intermediate metallic phase between the SC and I in 2D.

We estimate the critical $V_c$ and the corresponding $\rho_{dc}(V_c)$ from the raw data presented above, and similar data at other $|U|$ by 3 different methods. The first estimate of $V_c$ is obtained from measurements of $D_s$ at temperatures low enough that the pairing correlations are well formed across the entire lattice. Since $D_s \simeq (V - V_c)^\zeta$ with [4] $\zeta = z\nu > 1$, we expect $V_c$ to lie in the small $D_s$ tail in Fig. 1(b). The second estimate comes from the higher temperature crossing plots like Fig. 2(b), and the final estimate comes from the quality of the Drude fit, which works only for $V \simeq V_c$, as described above. On the 8x8 systems studied, we find excellent agreement between the three methods, as shown in Figs. 2(a) and 3(a). However, at this point we cannot rule out the effect of finite size corrections on our results [13].

We find evidence for only two phases in Fig. 3: a superfluid phase, with $D_s > 0$ and $\rho_{dc} = 0$ at $T = 0$, and an insulating state with $D_s = 0$ and $\rho_{dc} = \infty$ at $T = 0$. We do not find any clear evidence in our numerics for a metallic phase ($D_s = 0$ and $\rho_{dc} =$ finite at $T = 0$). For $|U| = 0$ we know from the scaling theory of Anderson localization that there is no metallic phase in 2D [2] (although this would be impossible to check for low disorder on the small systems under study here). In the opposite extreme $|U| \to \infty$ one obtains (short range) repulsive bosons in a random potential. There are general arguments [14] against the existence of a Bose metal with short range interactions at $T = 0$ in any dimension. Thus, if there is a metallic phase in the 2D model we study, it can only exist in the intermediate $|U|$ regime, where the QMC should be most reliable. Note that the phase diagram for the same model (1) in 3D would certainly be expected to have a metallic phase for small $|U|$.

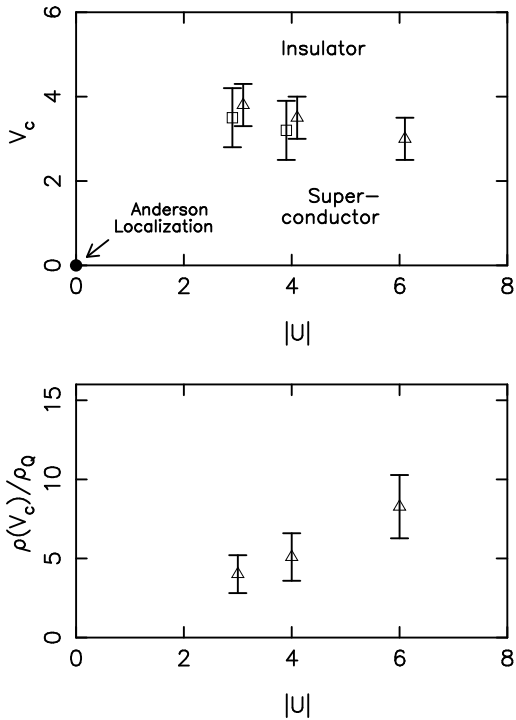

FIG. 3. (a): Estimates for $V_c$ at intermediate couplings $|U| = 3, 4, 6$ from vanishing of $D_s$ (squares), and from $\rho_{dc}$ crossing (triangles), which have been offset for clarity. The full circle at the origin is the $U = 0$ result $V_c = 0$ for 2D non-interacting electrons. We expect $V_c \propto t^2/U$ at large couplings and $V_c \sim t \exp(-t/|U|)$ at small U.
(b): Estimates for $\rho_{dc}(V_c)$ for $|U| = 3, 4, 6$ obtained from the $\rho_{dc}$ crossing plots.

While the results of Fig. 2(a,b) are in many ways reminiscent of experiments on the resistivity of thin films [1], we do not observe a "universal resistivity" [16] in the sense that $\rho_{dc}$ shows considerable variation with $|U|$. Recently, in a set of experiments on MoGe films, [1](d) a similar observation of sample dependent resistance was reported. In our model, $|U|$ directly controls the contribution of the unpaired electrons to the resistivity relative to the pairs. Our results for $\rho_{dc}$ increasing with $|U|$ are consistent with a universal boson or pair resistivity and a fermionic contribution that increases with $|U|$.

We gratefully acknowledge the hospitality of the Institute of Theoretical Physics, Santa Barbara, where this work was started. We thank K. Runge for providing useful scripts for doing the disorder averaging. The numerical calculations were performed primarily on Cray C-90 computers at SDSC and NERSC. This work was supported by the NSF under grants No. DMR92-06023 and No. ASC-9405041 (R.T.S.), and by DOE grant No. W-31-109-ENG-38 (N.T. and M.R.). R.T.S. thanks the hospitality of Argonne National Laboratory.